\documentstyle[aps,prl,multicol,epsf]{revtex}
\begin{document}
\draft

\title{Correlated sampling in quantum Monte Carlo: a route to forces}

\author{Claudia Filippi}
\address{Department of Physics, National University of Ireland, Cork, Ireland}
\author{C. J. Umrigar}
\address{Cornell Theory Center and Laboratory of Atomic and Solid State
Physics, Cornell University, Ithaca, New York 14853}
\date{\today}
\maketitle
\begin{abstract}
In order to find the equilibrium geometries of molecules and solids
and to perform ab initio molecular dynamics,
it is necessary to calculate the forces on the nuclei. We present a
correlated sampling method to efficiently calculate numerical forces and
potential energy surfaces  in diffusion Monte Carlo.
It employs a novel coordinate transformation, earlier used in variational Monte Carlo,
to greatly reduce the statistical error.
Results are presented for first-row diatomic molecules.

\end{abstract}

\begin{multicols}{2}
\setcounter{collectmore}{5}

Over the past decade, quantum Monte Carlo (QMC)
methods~\cite{general,UWW,FU} have been used to calculate
the structural and electronic properties of a variety of atoms,  clusters
and solids. For systems with large numbers of electrons, QMC methods at
present provide the most accurate benchmark calculations of structural
energies.
However, a major difficulty of QMC methods has been the determination of
equilibrium geometries and potential energy surfaces.  Hence, most QMC calculations
have been performed on geometries obtained with either density functional
theory (DFT) or conventional quantum chemistry methods.
The computation of forces on nuclei has been a stumbling block
that has limited a more widespread use of QMC methods.

DFT methods or standard quantum chemistry techniques use the Hellman-Feynman
theorem to compute the forces on nuclei~\cite{HF}.
Unfortunately, this is not practical within QMC for three reasons. First, the
wave functions used in QMC calculations are usually not obtained by minimizing
the energy.  Therefore, if Hellman-Feynman theorem were employed in variational
Monte Carlo (VMC), the forces would have a systematic error.
Second, in fixed-node diffusion Monte Carlo (DMC) the Hellman-Feynman force
has an error due to the discontinuity in the derivative of the fixed-node
wave function at nodes~\cite{HNR}.
Finally, in both VMC and DMC, the statistical fluctuations would be too large
since the fluctuations of the potential energy are
much larger than those of the total energy.

Alternatively, one could simply compute energy differences to obtain either
forces (for an infinitesimal displacement of the ions) or the full potential
energy surface of the system.  However, while quantum chemistry methods can
rely on having an approximately constant and smoothly varying error in the
energy, a major disadvantage of QMC methods is
that, in addition to systematic errors, one has
statistical errors which make the determination of energy
differences or smooth potential energy surfaces very expensive in computer
time.
Even though it is not possible to entirely eliminate the statistical
errors, it is possible, by using correlated
sampling~\cite{KalosWhitlock}, to make the statistical errors
in the relative energies of different geometries much smaller than the
errors in the separate energies and to make them vanish in the limit that the
two geometries become identical.
In the past, the correlated sampling technique has been used within
VMC~\cite{ColdwellHe2,Sanibel}
but there have been
very few attempts~\cite{CorrSampDMC} to
extend the approach to DMC, and these were approximate and/or inefficient and
were tested only on H$_2$, H$_3^+$ and LiH.

In this paper, we present a novel DMC correlated sampling technique
to efficiently compute accurate forces and potential energy surfaces.
The DMC bond lengths of first-row diatomic molecules computed with this algorithm
are found to be in better agreement with experiment values than are the VMC,
Hartree-Fock (HF),
local density approximation (LDA) and generalised gradient approximation (GGA) values.

{\it Correlated sampling in variational Monte Carlo}.
Instead of performing independent VMC runs which
would have independent statistical errors, one generates MC configurations
for a reference situation only. The MC configurations are sampled from
$\psi^2$ where $\psi$ is the wave function for the reference situation.
Then, unbiased expectation values for somewhat
different secondary wave functions $\psi_{\rm s}$ are obtained by reweighting the configurations
sampled from $\psi^2$, e.g., for Hamiltonians ${\cal H}$ and ${\cal H}_{\rm s}$,
\begin{eqnarray}
E_{\rm s}-E &=& \frac{\langle\psi_{\rm s}\vert{\cal H}_{\rm s}\vert\psi_{\rm s}\rangle}
{\langle\psi_{\rm s}\vert\psi_{\rm s}\rangle}-\frac{\langle\psi\vert{\cal H}\vert\psi\rangle}
{\langle \psi \vert \psi \rangle}\nonumber\\
&=& {1 \over N_{\rm conf}}\sum_{i=1}^{N_{\rm conf}} \left\{\frac{{\cal H}_{\rm s}\psi_{\rm s}({\bf R}_i)}
{\psi_{\rm s}({\bf R}_i)} W_i-\frac{{\cal H}\psi({\bf R}_i)}{\psi({\bf R}_i)}\right\},
\label{vmc_samp}
\end{eqnarray}
where the weights of the $N_{\rm conf}$ MC configurations are
\begin{eqnarray}
W_i=\frac{N_{\rm conf} \left\vert\psi_{\rm s}({\bf R}_i)/\psi({\bf R}_i)\right\vert^2}
{\sum_{i=1}^{N_{\rm conf}} \left\vert\psi_{\rm s}({\bf R}_i)/\psi({\bf R}_i)\right \vert^2},
\label{vmc_weights}
\end{eqnarray}
and ${\bf R} \equiv ({\bf r}_1,\ldots,{\bf r}_N)$.
The statistical error in $E_{\rm s}-E$ is considerably smaller than that in
$E_{\rm s}$ or $E$, making this method of calculating $E_{\rm s}-E$
more efficient than performing independent VMC computations of $E_{\rm s}$ and $E$.

{\it Space-warp coordinate transformation}.
Since we want to compute the relative energy of two different geometries, the
Hamiltonians ${\cal H}$ and ${\cal H}_{\rm s}$ in Eq.~\ref{vmc_samp} correspond to
two sets of nuclear coordinates, ${\bf R}_\alpha$ and ${\bf R}_\alpha^{\rm s}$,
for the reference and the secondary geometry, respectively.

The electronic coordinates sampled from the reference wave function
$\psi^2$ will not be optimal for computing the energy $E_{\rm s}$ corresponding
to the nuclear coordinates ${\bf R}_\alpha^{\rm s}$, since the electron density
will be peaked at ${\bf R}_\alpha$ rather than at ${\bf R}_\alpha^{\rm s}$.
To solve this problem, a mapping of the electron coordinates
was introduced in Ref.~\cite{Sanibel}
such that those electrons that are close to a given nucleus move almost
rigidly with that nucleus:
\begin{eqnarray}
{\bf r}^{\rm s}_i =  {\bf r}_i + \sum_{\alpha =1}^{N_{\rm atoms}}
({\bf r}_\alpha^{\rm s}- {\bf r}_\alpha) \; \omega_\alpha({\bf r}_i),\label{warp}
\end{eqnarray}
where
\begin{eqnarray}
\omega_{\alpha}({\bf r}_i) = \frac{F(\vert\, {\bf r}_i - {\bf r}_\alpha \vert)}
{\sum_{\beta =1}^{N_{\rm atoms}} F(\vert\, {\bf r}_i - {\bf r}_\beta \vert)};
\;\;\; \sum_{\alpha =1}^{N_{\rm atoms}} \omega_\alpha({\bf r}_i) =1\,.
\end{eqnarray}
(We use Latin indices for electronic coordinates and Greek indices for nuclear
coordinates.)
$F(r)$ is any sufficiently rapidly decaying function such as
$r^{-\kappa}$, $\exp(-\kappa r)$ or $\exp(\kappa/r)$.
The reduction in statistical error due to the use of these
{\it warped} electronic coordinates for the secondary geometries
is large and almost independent of the choice for $F(r)$
and $\kappa$.  Unless otherwise stated, all results in this paper
were computed with $F(r)=r^{-\kappa}$ and $\kappa=4$.

The equation for $E_{\rm s}-E$ (Eq.~\ref{vmc_samp}) is now
\begin{eqnarray}
E_{\rm s}-E = {1 \over N_{\rm conf}} \sum_{i=1}^{N_{\rm conf}} \left(
\frac{{\cal H}_{\rm s}\psi_{\rm s}({\bf R}^{\rm s}_i)}{\psi_{\rm s}({\bf R}^{\rm s}_i)} W_i
- \frac{{\cal H}\psi({\bf R}_i)}{\psi({\bf R}_i)}\right),
\label{vmc_warp}
\end{eqnarray}
where
\begin{eqnarray}
W_i=\frac{N_{\rm conf} \left \vert\psi_{\rm s}({\bf R}^{\rm s}_i) / \psi({\bf R}_i)\right \vert^2 J({\bf R}_i)}
{\sum_{j=1}^{N_{\rm conf}} \left \vert\psi_{\rm s}({\bf R}^{\rm s}_j) / \psi({\bf R}_j) \right \vert^2 J({\bf R}_j)}
\,,
\label{weight_warp}
\end{eqnarray}
and $J({\bf R})$ is the Jacobian for the transformation (Eq.~\ref{warp}).

{\it Correlated sampling in diffusion Monte Carlo}.
In DMC, the imaginary-time evolution operator $\exp(-{\cal H}\tau)$ is used to
project out the ground state from the trial wave function within the fixed-node
and the short-time approximations~\cite{dmc}. The primary walk is generated
according to the stochastic implementation of the integral equation:
\begin{eqnarray}
f({\bf R}',t+\tau)=\int d{\bf R}\, G({\bf R}',{\bf R},\tau)\, f({\bf R},t)\,,
\end{eqnarray}
where $G({\bf R}',{\bf R},\tau)=\left<{\bf R}'|\exp\{-{\cal H}\tau\}|{\bf R}\right>$.
If we introduce importance sampling using the reference wave function $\psi$,
the importance sampled Green's function for small values of $\tau$ (short-time
approximation) is given by the product of three factors, diffusion,
drift and growth/decay:
\begin{eqnarray}
\tilde{G}({\bf R}',{\bf R},\tau)&=&\frac{1}{(2\pi\tau)^{\frac{3N}{2}}}
e^{-\frac{({\bf R}'-{\bf R}-{\bf V}({\bf R})\tau)^2}{2\tau}}
e^{S({\bf R}',{\bf R},\tau)}
,\label{green}
\end{eqnarray}
where ${\bf V}=\nabla \psi({\bf R})/\psi({\bf R})$ and
$S({\bf R}',{\bf R},\tau)=(E_{\rm T}-E_{\rm L}({\bf R}')-E_{\rm L}({\bf R}))\tau/2$
with $E_{\rm L} ={\cal H}\psi({\bf R})/\psi({\bf R})$.
At time $t$, a set of primary walkers characterized by the pairs $({\bf R}_i,w_i)$
is a random realization of the distribution $f$:
\begin{eqnarray}
f({\bf R},t)=\sum_i w_i \; \delta({\bf R}-{\bf R}_i).
\end{eqnarray}
Each walker executes a branching random walk: a walker originally at ${\bf R}$
drifts to ${\bf R}+{\bf V}({\bf R})\tau$ and then diffuses to ${\bf R}'$ according
to the Gaussian term in Eq.~\ref{green}.
To ensure that, in the limit of perfect importance sampling (i.e. $\psi$ is
the ground state wave function $\psi_0$), we are sampling $\psi^2$ despite the
short-time approximation in the Green's function, the move is accepted with
probability
\begin{eqnarray}
p=\min\left\{1,\frac{|\psi({\bf R}')|^2 \; \tilde{T}({\bf R},{\bf R}',\tau)}
{|\psi({\bf R})|^2 \; \tilde{T}({\bf R}',{\bf R},\tau)}\right\},
\label{accept}
\end{eqnarray}
as prescribed by the detailed balance condition. We denote by $\tilde{T}$
the drift-diffusion part of the Green's function $\tilde{G}$.
Finally, the weight of the walker is multiplied by $\exp[S({\bf R}',{\bf R},\tau)]$.
In practice, we employ the modified version of $\tilde{G}$ presented in
Ref.~\cite{UNR} that takes into account the non-analyticities of ${\bf V}$
and $E_{\rm L}$ at the nodes and particle coalescence points.

Given a primary walk generated according to Eq.~\ref{green}, the
secondary walk is specified by the space-warp transformation
(Eq.~\ref{warp}).
Two complications, absent in VMC, arise for correlated sampling in DMC.
First of all, the dynamics of the secondary walker should have been governed by an
importance sampled Green's function constructed from the secondary wave
function $\psi_{\rm s}$, $\tilde{G}_{\rm s}({\bf R}^{\rm s}{}',{\bf R}^{\rm s},\tau)$,
and the move should have been accepted with probability
\begin{eqnarray}
p_{\rm s}=\min\left\{1,\frac{|\psi_{\rm s}({\bf R}^{\rm s}{}')|^2
\; \tilde{T}_{\rm s}({\bf R}^{\rm s},{\bf R}^{\rm s}{}',\tau)}{|\psi_{\rm s}({\bf R}^{\rm s})|^2
\; \tilde{T}_{\rm s}({\bf R}^{\rm s}{}',{\bf R}^{\rm s},\tau)}\right\}.
\end{eqnarray}
However, the secondary-geometry move was effectively proposed according to the
drift-diffusion Green's function $\tilde{T}({\bf R}',{\bf R},\tau)/J({\bf R}')$
and accepted with probabilty $p$ defined in Eq.~\ref{accept}.
To correct for the wrong dynamics,
we should multiply the weights of the secondary walkers by
\begin{eqnarray}
r\,\frac{\tilde{G}_{\rm s}({\bf R}^{\rm s}{}',{\bf R}^{\rm s},\tau)}
{\tilde{T}({\bf R}',{\bf R},\tau)/J({\bf R}')}\,\,,
\label{correction}
\end{eqnarray}
where $r=p_{\rm s}/p$ if the move is accepted and $r=(1-p_{\rm s})/(1-p)$ if
the move is rejected.
However, these products fluctuate wildly ($r$ can be
anywhere between zero and infinity).
Therefore, it is not practical to follow this route to perform correlated
sampling unless bounds can be  placed on the ratios while at the same time
ensuring that unbiased results are obtained in the $\tau \to 0$ limit.

An additional complication is the common practice in fixed-node DMC to
reject  moves that cross nodes. If a primary walker attempts to cross,
the move is rejected, $p$ in Eq.~\ref{accept} is set to zero, and the
ratio $r$ for the secondary walker (Eq.~\ref{correction}) becomes ill-defined.
Moreover, if primary and secondary walkers were to be treated on the same
footing ($p_{\rm s}$ set to zero when the secondary walker crosses its
own nodes), the weights of the secondary walkers would all
become zero in a sufficiently long run.
Even though this problem can be easily overcome
since it is legitimate to do fixed-node DMC allowing walkers to cross
nodes~\cite{UNR}, reweighting as in Eq.~\ref{correction} remains
impractical due to the large fluctuations.

In this paper, we propose an alternative correlated sampling algorithm. Our
algorithm is approximate but very accurate. Given the successful implementation
of correlated sampling within VMC, we wish to devise a scheme that differs
as little as possible from VMC but yields results very close to the fixed-node DMC
limit for the secondary geometries. We perform the primary walk as described above
and generate the secondary walks according to the space-warp transformation. In the
averages, we retain the ratios of the secondary and primary wave functions as
would be done in VMC (Eqs.~\ref{vmc_warp} and~\ref{weight_warp}).
The secondary weights are the primary ones multiplied by the product of the factors
$\exp[S_{\rm s}({\bf R}^{\rm s}{}',{\bf R}^{\rm s},\tau_s)-S({\bf R}{}',{\bf R},\tau)]$
for the last $N_{\rm proj}$ generations.
Note that, in the exponential factors, we introduced $\tau_{\rm s}$, a time-step for
the secondary path, in general different from the time-step $\tau$ used for the
primary walk.
Due to the warp transformation, the
secondary moves were effectively proposed with a different time-step, $\tau_s$, in the
drift-diffusion term.
A sensible definition of $\tau_{\rm s}$ is
\begin{eqnarray}
\tau_{\rm s}=\tau \frac{\left<\Delta R_{\rm s}^2\right>}{\left<\Delta R^2\right>},
\label{tau_s}
\end{eqnarray}
where $\Delta R$ is the displacement resulting from diffusion,
and $\Delta R_{\rm s}$ is the displacement needed to take the secondary walker from
its drifted position to the position specified by the space-warp transformation.
Having computed $\tau_{\rm s}$ over the first equilibration blocks of our DMC
run, we will use this time-step for computing the drift and reweighting of the
secondary walk.
Therefore, each secondary geometry has a different time-step $\tau_{\rm s}$.

{\it Secondary geometry wave functions}.
We considered three choices for secondary geometry wave functions:\\
(1) The secondary geometry wave functions have the same parameters $\{\bf p\}$ as
the primary one but the coordinates are relative to the new nuclear
positions: $\psi_{\rm s}({\bf R}_i,{\bf R}_\alpha^{\rm s})= 
\psi({\bf R}_i,{\bf R}_\alpha^{\rm s},{\bf p}_{\rm s})$ with ${\bf p}_{\rm s}={\bf p}$, 
possibly with the minimal changes required to impose the cusp conditions.\\
(2) The secondary geometry wave functions at warped electron positions are
related to the primary ones at the original positions,
$\psi_{\rm s}({\bf R}_i^{\rm s},{\bf R}_\alpha^{\rm s})={\psi({\bf R}_i,{\bf R}_\alpha,{\bf p})/
\sqrt{J({\bf R}_i)}}$. This wave function depends on the transformation (it was
used in Ref.~\cite[{\it b-c}]{CorrSampDMC} with a different transformation) and has
the advantage that the weights $W_i$ in (Eq.~\ref{vmc_weights}) are unity.
Surprisingly, it gives larger fluctuations of the energy differences than
choice (1).\\
(3) $\psi_{\rm s}({\bf R}_i,{\bf R}_\alpha^{\rm s}) = \psi({\bf R}_i,{\bf R}_\alpha^{\rm s},{\bf p}_{\rm s})$
with reoptimized parameters ${\bf p}_{\rm s}$. This choice gives the smallest fluctuation
of the energy differences and the best potential energy surface.\\
We calculate all molecules with choice (1) but also demonstrate the
superiority of choice (3) for B$_2$.

{\it Results and conclusions.}
The algorithms presented in the previous sections are tested on first-row homonuclear
dimers.  The primary wave functions~\cite{FU} were optimized close to the experimental bond
length by the variance minimization method~\cite{UWW}.
The potential energy curves were obtained with correlated sampling
from ten geometries, using the warp transformation
and recentered secondary geometry wavefunctions (choice (1) above).

\vspace*{-.4cm}
\noindent
\begin{minipage}{3.375in}
\begin{figure}
\centerline{\epsfxsize=8.0 cm \epsfysize=6.5 cm \epsfbox{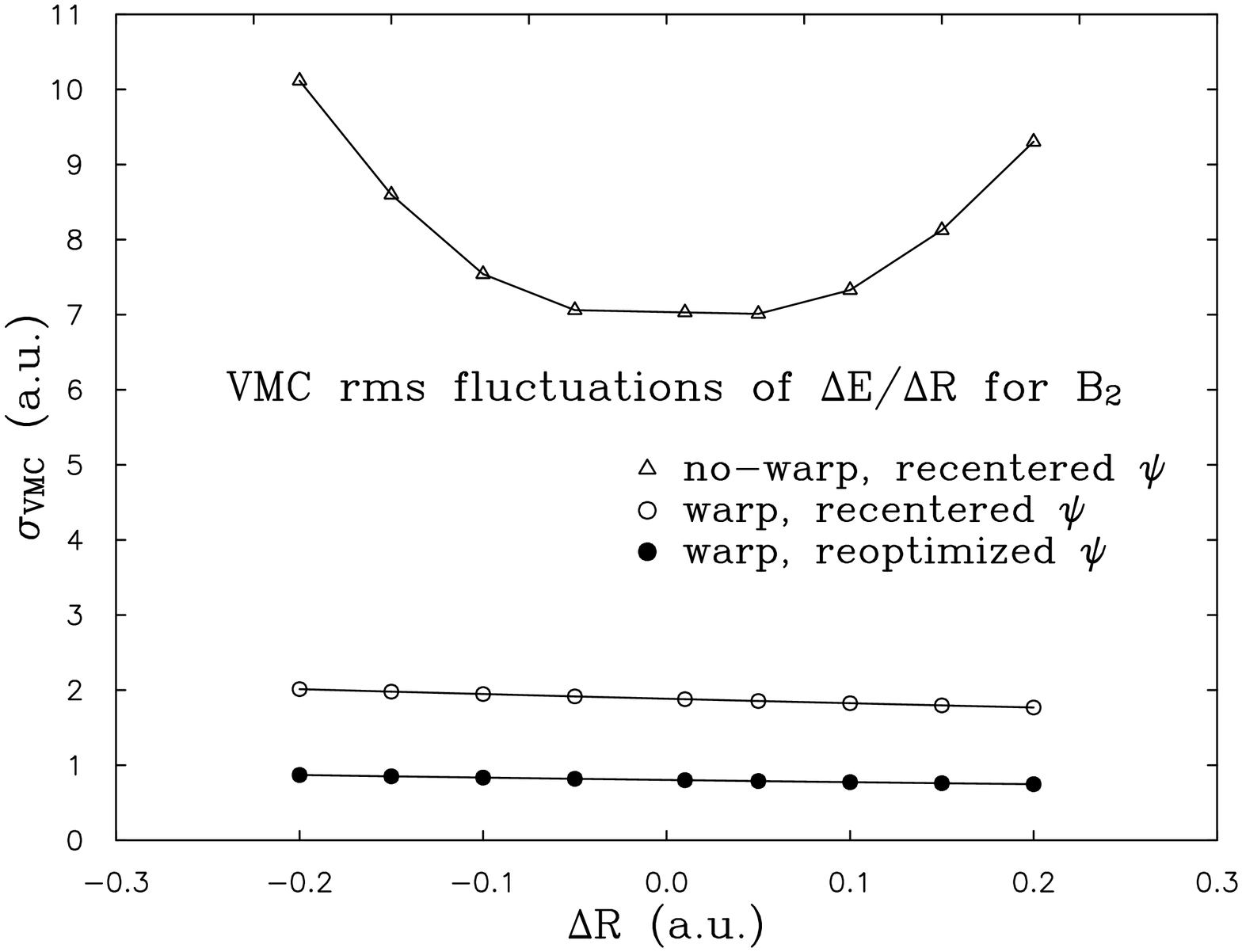}}
\vspace{.2cm}
\caption[]{VMC fluctuations ($\sigma_{\rm VMC}$) of the relative
energy of the primary and secondary geometries divided by the bond stretch
for B$_2$.
The smallest $\sigma_{\rm VMC}$ is achieved by using warping along with
reoptimized secondary wave functions.
}
\vspace{.2cm}
\label{fig1}
\end{figure}
\end{minipage}

To ascertain the efficiency of our method, we performed two additional
calculations for B$_2$; in the first, we omitted the warp transformation, whereas in the second
we employed reoptimized, rather than recentered, secondary wavefunctions.
In Fig.~\ref{fig1}, we present the VMC root-mean-square fluctuations ($\sigma_{\rm VMC}$)
of the relative energy of primary and secondary geometries divided by the atomic
displacement, $\Delta E / \Delta R$,  for B$_2$.
Introducing the warp transformation yields a reduction of about
a factor of 3.5-5 in $\sigma_{\rm VMC}$, which corresponds to a factor of 12-25
saving in computer time. Moreover, $\sigma_{\rm VMC}$ is only slightly dependent on
the secondary geometry used. As expected,  a further reduction in $\sigma_{\rm VMC}$
is obtained when the space-warp transformation is used in combination with
reoptimized, rather than recentered, secondary geometry wave functions.
The space-warp transformation was found to be of even greater help for heavier molecules,
e.g. for F$_2$ the reduction in the fluctuations was a factor of 3.2-8.2.

To test the accuracy of our DMC correlated sampling algorithm, we performed DMC
runs for H$_2$ and B$_2$ for three different primary geometries, (a) the equilibrium
geometry, (b) a geometry stretched by $0.2$ and (c) by $-0.2$ a.u.
The runs (a), (b) and (c) should
give identical potential energy curves if the algorithm were exact.
In Fig.~\ref{fig2}, we show results for B$_2$ that reveal the high accuracy of our
DMC algorithm: the three DMC curves are very close and clearly distinguishable from the
VMC results. These results are confirmed by the calculations for H$_2$ where, despite 
the use of an intentionally poor wave function, the three
curves gave the equilibrium bond lengths (a) 1.4014(2) (b) 1.4014(2) and
(c) 1.4015(2) a.u. The true equilibrium bond length, from a careful fit to
the
results of Ref.~\cite{Wolniewicz93}, is 1.4011 a.u.

\vspace*{-.4cm}
\noindent
\begin{minipage}{3.375in}
\begin{figure}
\centerline{\epsfxsize=8.0 cm \epsfysize=6.5 cm \epsfbox{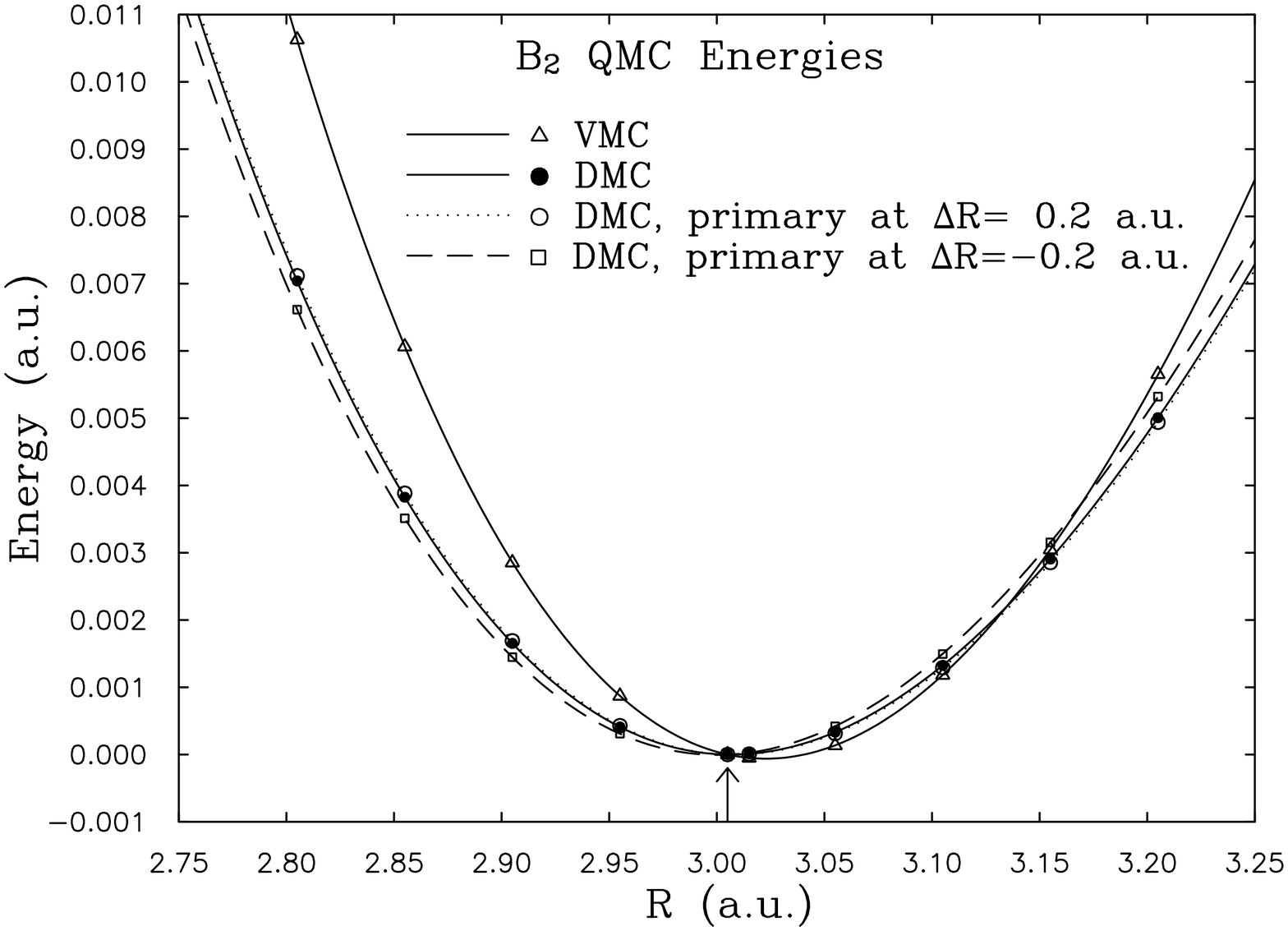}}
\vspace{.2cm}
\caption[]{Potential energy curve for B$_2$ in VMC and DMC. The three DMC curves
are obtained with three different primary geometries (equilibrium, stretched by 0.2 and
$-$0.2 a.u.) and using recentered wave functions. All curves are shifted with the energy
at the equilibrium distance (arrow) defined as the zero. Atomic units are used.}
\vspace{.2cm}
\label{fig2}
\end{figure}
\end{minipage}

To test the improvement resulting from employing $\tau_s$\,$\ne$\,$\tau$
(Eq.~\ref{tau_s}), we performed, for H$_2$, DMC correlated sampling with $\tau_{\rm s}$\,=\,$\tau$.
Since $\tau_{\rm s}$\,$>$\,$\tau$ for $\Delta R$\,$>$\,0 and $\tau_{\rm s}$\,$<$\,$\tau$ for
$\Delta R$\,$<$0,
we expect this potential energy curve to yield an equilibrium bond-length that is too short.
The equilibrium bond length is indeed 1.4003(2) a.u., which is 4 standard deviations
from the true bond-length, whereas that obtained with our $\tau_s$\,$\ne$\,$\tau$ algorithm,
1.4014(2) a.u., is 1.5 standard deviation from the true bond-length.

Having ascertained the accuracy and efficiency of our algorithm, we computed the bond lengths of all
first-row dimers with VMC and DMC correlated sampling. In Table~\ref{tab1}, we list the
errors in the bond lengths obtained from restricted Hartree-Fock (RHF)~\cite{R_HF},
LDA~\cite{R_LDA}, GGA~\cite{R_GGA}, VMC and DMC.
The RHF results show the worst agreement with experiment, with Be$_2$ not being bound.
The DMC errors are, in all cases, either smaller than or comparable
to those from VMC, and are smaller than LDA and GGA errors by a factor of
3.9 and 2.6, respectively.

In this letter, we presented an efficient method to compute numerical forces in DMC,
a long-standing unsolved problem in QMC techniques.
The method is very accurate and was tested on
first-row dimers, where the DMC bond lengths were found to agree with experiment
better than those from HF, LDA, GGA and VMC.

\noindent
\begin{minipage}{3.375in}
\begin{table}[htb]
\caption[]{Experimental~\cite{Herzberg} bond lengths (in a.u.) of first-row dimers
and theoretical errors in RHF, LDA, GGA, VMC and DMC.
}
\label{tab1}
\begin{tabular}{lcccccc}
{molecule}  & {Expt.} & {RHF} & {LDA} & {GGA} & {VMC} & {DMC} \\[.1cm]
\hline\\[-.2cm]
Li$_2$ & 5.051 &  0.270 &  0.069 &  0.057  &  0.101(2)  &  0.018(3)\\
Be$_2$ & 4.630 &   --   & -0.109 & -0.001  & -0.069(3)  & -0.014(5)\\
B$_2$  & 3.005 &  0.086 &  0.025 &  0.042  &  0.018(2)  &  0.002(2)\\
C$_2$  & 2.348 & -0.007 &  0.006 &  0.023  &  0.006(2)  &  0.008(1)\\
N$_2$  & 2.074 & -0.061 & -0.006 &  0.011  &  0.012(2)  &  0.007(1)\\
O$_2$  & 2.282 & -0.107 & -0.012 &  0.044  &  0.028(2)  &  0.023(4)\\
F$_2$  & 2.668 & -0.161 & -0.053 &  0.040  &  0.021(4)  &  0.015(5)\\
rms    &  -- & $\infty$ &  0.054 &  0.036  &  0.049     &  0.014 \\
\end{tabular}
\end{table}
\end{minipage}

{\it Acknowledgements.}
We thank S. Fahy for useful discussions. This work was begun during a visit to
the Institute for Nuclear Theory in Seattle and is funded by Sandia National Laboratory.

\end{multicols}

\end{document}